\newcommand\aj{AJ}%
\newcommand\araa{ARAA}%
\newcommand\apj{ApJ}%
\newcommand\aap{A \& A}%
\newcommand\pasp{PASP}%
\newcommand\mnras{MNRAS}
\newcommand\nat{Nature}

\newcommand\lesssi{\mathrel{\hbox{\rlap{\hbox{\lower4pt\hbox{$\sim$}}}\hbox{$<$}}}}
\newcommand\gtrsi{\mathrel{\hbox{\rlap{\hbox{\lower4pt\hbox{$\sim$}}}\hbox{$>$}}}}

\documentclass[
    ,final            
  ]
  {aipproc}

\layoutstyle{6x9}

\begin{document}

\title{A Spectropolarimetric Comparison of the Type II-Plateau Supernovae SN 2008bk
  and SN 2004dj}


\author{Douglas C. Leonard}{
  address={Department of Astronomy, San Diego State University, San Diego, CA
  92182, USA; leonard@sciences.sdsu.edu} }

\author{Luc Dessart}{
  address={Laboratoire d'Astrophysique de Marseille, Universit\'{e} de Provence,
CNRS, 38 rue Fr\'{e}d\'{e}ric-Joliot-Curie, F-13388 Marseille Cedex 13, France} }

\author{D. John Hillier}{
  address={Department of Physics and Astronomy, University of Pittsburgh, 3941
  O'Hara Street, Pittsburgh, PA 15260, USA} }

\author{Giuliano Pignata}{
  address={Departamento de Ciencias Fisicas, Universidad Andres Bello,
  Avda. Republica 252, Santiago, Chile} 
  ,altaddress={Departamento de Astronomia, Universidad de Chile, Casilla 36-D,
  Santiago, Chile} 
}

\begin{abstract}

The Type II-Plateau supernova (SN~II-P) SN~2004dj was the first SN~II-P for
which spectropolarimetry data were obtained with fine temporal sampling before,
during, and after its fall off of the photometric plateau -- the point that
marks the transition from the photospheric to the nebular phase in SNe~II-P.
Unpolarized during the plateau, SN~2004dj showed a dramatic spike in
polarization during the descent off of the plateau, and then exhibited a smooth
polarization decline over the next two hundred days.  This behavior was
interpreted by Leonard et al. (2006) as evidence for a strongly non-spherical
explosion mechanism that had imprinted asphericity only in the innermost
ejecta.  In this brief report, we compare nine similarly well-sampled epochs of
spectropolarimetry of the Type II-P SN 2008bk to those of SN~2004dj.  In
contrast to SN~2004dj, SN~2008bk became polarized well before the end of the
plateau and also retained a nearly constant level of polarization through the
early nebular phase.  Curiously, although the onset and persistence of
polarization differ between the two objects, the detailed spectropolarimetric
characteristics at the epochs of recorded maximum polarization for the two
objects are extremely similar, feature by feature.  We briefly interpret the
data in light of non-Local-Thermodynamic Equilibrium, time-dependent
radiative-transfer simulations specifically crafted for SN II-P ejecta.

\end{abstract}

\maketitle

\vspace*{-0.4in}

\section{Introduction}
\vspace*{-0.1in}

Spectropolarimetry offers a direct probe of early-time SN geometry since a hot,
young SN atmosphere is dominated by electron scattering, which by its nature is
highly polarizing.  For an unresolved source that has a spherical distribution
of scattering electrons, the directional components of the electric vectors of
the scattered photons cancel exactly, yielding zero net linear polarization.
Any asymmetry in the distribution of the scattering electrons, or of absorbing
material overlying the electron-scattering atmosphere, results in incomplete
cancellation, and produces a net polarization
\citep[see, e.g.,][]{Leonard11b}.  

Type II-Plateau supernovae (SNe~II-P) are the canonical core-collapse events
that arise from isolated red supergiant stars \citep{Smartt09a}.  The precise
nature of the mechanism responsible for the stellar explosion, however, remains
the subject of considerable debate \cite{Nordhaus10}.  One key diagnostic is
explosion geometry.  A global asphericity of SN ejecta will manifest itself
through two main spectropolarimetric signatures.  First, there will be
significant polarization in spectral regions not dominated by line opacity.
For SNe~II-P, this includes the broad spectral region 6800--8200 \AA.  Strong
emission lines should generally be unpolarized, since directional information
is lost as photons are absorbed and reemitted in a line \citep{Hoflich96}
\citep[see, however,][]{Dessart11}. Finally, polarization increases are
anticipated in the troughs of strong P-Cygni absorption lines.  This results
primarily from selective blocking of more forward-scattered and, hence, less
polarized, light in the trough regions.  A critical point to bear in mind is
that the longer one waits after the explosion, the deeper into the ejecta one
can see.  For SNe II-P, the story unfolds over an extended time period as the
photosphere recedes (in mass coordinates) back through the ejecta as the
hydrogen recombination front progresses from the outside inwards through the
thick hydrogen envelope. The transition from the opaque, ``photospheric'' phase
to the transparent, ``nebular phase'' is signaled by the sudden drop in
luminosity that marks the end of the photometric plateau.

In initial studies, SNe~II-P were found to have minimal intrinsic polarizations
at the generally early times at which they were observed.  Low-contrast line
features were sometimes seen in the data, but no obvious strong continuum
polarization was found \citep[e.g., ][]{Leonard3,Leonard4}; for a recent review
of SN spectropolarimetry of all types, see \cite{Wang08}. The situation
changed dramatically with observations of SN~2004dj, however, for which a
sudden ``spike'' in polarization was seen during the steepest part of the
descent off of the plateau \citep{Leonard13}. Using existing models
\cite{Hoflich91}, it was concluded that the late
onset of the polarization suggested a fundamentally non-spherical explosion,
with the asphericity cloaked at early times by the massive, opaque, hydrogen
envelope.

An obvious observational question presented itself: Was SN~2004dj a unique (or
rare) event?  Or, would other similarly well-observed SNe~II-P display
similar behavior?  \citet{Chornock10} and \citet{Leonard09} provide partial
answers from analysis of four additional SNe~II-P: Three (SNe 2006my, 2006ov,
and 2007aa) exhibit large, late-time (end of photometric plateau and beyond; )
polarization \citep{Chornock10}, while one (SN~2004et) exhibits strong and
temporally increasing polarization that began at least three months before the
end of its plateau phase \citep{Leonard09}.  Detailed comparison of these
datasets with that of SN~2004dj is hindered, however, by modest temporal
sampling or phase coverage.
\vspace*{-0.2in}

\section{Spectropolarimetry of SN 2008bk}
\vspace*{-0.1in}

Here we show initial results from our program to obtain critically sampled
spectropolarimetry of SNe~II-P with the European Southern Observatory Very
Large Telescope.  Figure~1 presents nine spectropolarimetric epochs for our
most well observed SN~II-P, SN~2008bk, along with the data of SN~2004dj
from \cite{Leonard13}.  Unlike the case for SN~2004dj, SN~2008bk was strongly
polarized well {\it before} the end of the plateau (much like SN~2004et), with
the level rising through the plateau.  Despite the different temporal behavior,
it is intriguing that at the point of maximum recorded polarization (i.e., day
31 with respect to the end of the plateau for SN~2004dj and day -21 for
SN~2008bk) the spectropolarimetric {\it signatures} of SN~2008bk are virtually
identical to those of SN~2004dj.  This similarity may suggest a common cause
--- or, at least, geometry.  Following the plateau drop-off, the polarization
of SN~2008bk remains remarkably constant.  In fact, although not shown in
Figure~1, the spectropolarimetry from days 9 through 120 post-plateau (epochs 5
through 9) are virtually indistinguishable from one another.  This is quite
different from the behavior of SN~2004dj.

\begin{figure}[h!]
  \includegraphics[height=0.5\textheight]{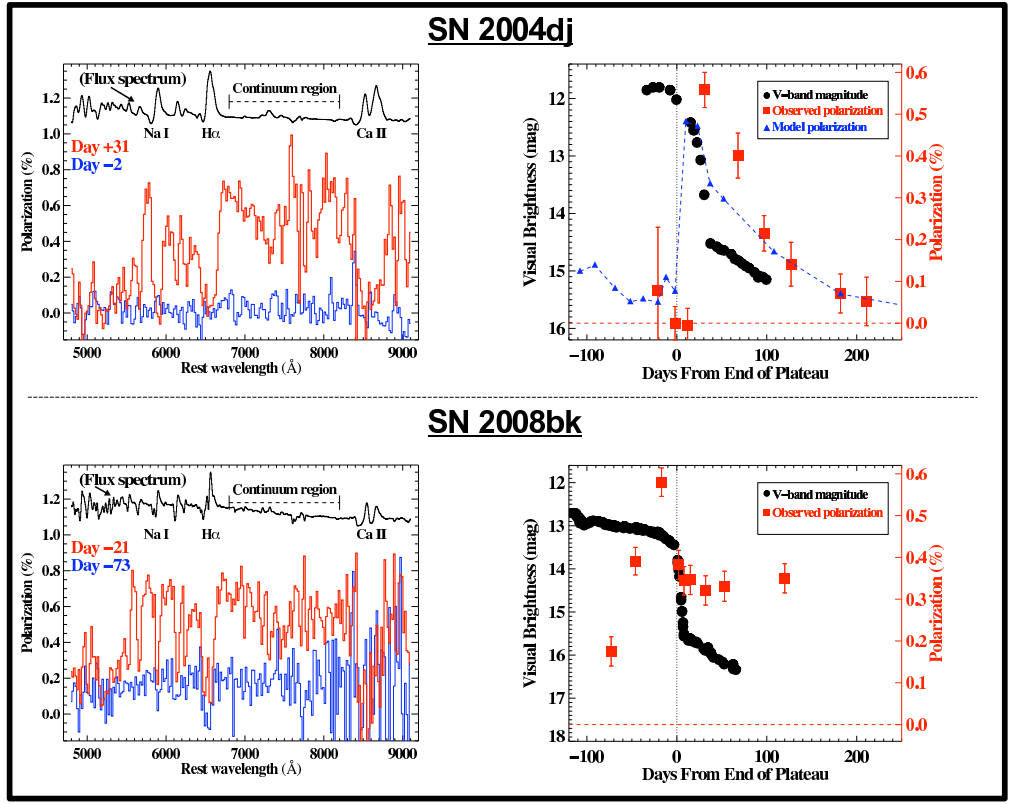} 
\caption{Comparison of the polarization of SN~2004dj with SN~2008bk.  {\bf
Top, left}: Linear polarization of SN~2004dj at two representative epochs with
respect to the end of the plateau phase (labels indicate the phase in days from
the end of the plateau, defined as the point where the light curve begins a
precipitous drop).  {\bf Top, right}: Continuum polarization (from the spectral
region 6800 \AA --- 8200 \AA) and $V$-band magnitude of SN~2004dj; the expected
polarization derived from the models of \cite{Dessart11} and discussed in the
text is also shown (filled triangles connected by a dashed line).  {\bf
Bottom:} Same as the top, but for SN~2008bk.  Note that an interstellar
polarization (ISP) of $\sim 0.29\% $ has been removed from the SN~2004dj
data \cite{Leonard13}, whereas no ISP has been removed from the SN~2008bk data,
because it is estimated to be $\lesssi 0.1\% $ from the depolarizations seen
across strong emission lines (e.g., H$\alpha$, Ca~II) at all epochs.}

\end{figure}

\vspace*{-0.4in}
\subsection{Theoretical Modeling}
\vspace*{-0.1in}

We have developed non-Local-Thermodynamic Equilibrium, time-dependent
radiative-transfer simulations specifically crafted for
SNe~II-P \cite{Dessart11}.  The simulations, based on the one-dimensional
SN~II-P models of \cite{Dessart11b}, compute the expected polarization (in both
lines and continuum) as a function of time for an asphericity generated by a
latitudinal density enhancement.  A significant finding is that temporal
polarization changes in SNe~II-P do not {\it necessarily} implicate a change in
the asymmetry of the ejecta.  For instance, in Figure~1 (top, right) we include
the expected temporal evolution of the continuum polarization for an SN~II-P
with an oblate equator-to-pole density enhancement of 1.67, viewed edge-on.
This model reproduces what is observed for SN~2004dj quite well, with the
sudden ``spike'' in polarization now explained not by a dramatic increase in
asphericity but, rather, by the sharp reduction in optical depth (i.e., fewer
electron scatters for outgoing photons) that occurs at the plateau's end.  The
post-plateau polarization decline is also reproduced as a consequence of the
expected $1/t^2$ dependence of the ejecta optical depth during the nebular
phase \cite{Leonard13,Dessart11}.

While models with a dominant and fixed large-scale asymmetry fit the
polarization behavior of SN~2004dj, the substantial early time polarization
seen in SN~2008bk (and SN~2004et) is not replicated; in fact, current models
with equator-to-pole density enhancements as extreme as 5.0 are unable to
produce significant (i.e., > 0.1\%) polarization during the
mid-plateau \cite{Dessart11}.  The persistent late-time polarization seen in
SN~2008bk also presents a modeling challenge.  

It is worth noting that, in addition to their spectropolarimetric differences,
SNe~2008bk, 2004dj, and 2004et also exhibit photometric and spectroscopic
differences \cite{Maguire10,Vandyk11}, some of which are clearly evident for
SN~2004dj and SN~2008bk in Figure~1, such as the rapidity of the plateau
drop-off.  These differences may be associated with explosion physics but might
also be related to the progenitor's core structure and the $^{56}{\rm
Ni}$-mixing properties (morphology, amount, depth below the hydrogen-rich
envelope; see \cite{Chugai06}) as well.  It is hoped that with the benefit of
further modeling we may begin to connect all of these parameters and develop a
consistent interpretation of the asymmetries indicated by the polarization of
SNe~II-P.



\vspace*{-0.15in}

\begin{theacknowledgments}
\vspace*{-0.1in}
We thank for Bruno Leibundgut, Stephane Blondin, Tom Matheson, Robert Kirshner,
Malcom Hicken, and Brian Schmidt for assistance with the observations.  DCL
acknowledges support from NSF grant AST-1009571, under which part of this
research was carried out.  Based on observations collected at the European
Organisation for Astronomical Research in the Southern Hemisphere, Chile, under
observing programs 081.D-0128, 082.D-0151, and 085.D-0391 (PI: Dessart).
\vspace*{-0.1in}

\end{theacknowledgments}



\bibliographystyle{aipproc}   


\vspace*{-0.1in}

\begin{thebibliography}{12}
\expandafter\ifx\csname natexlab\endcsname\relax\def\natexlab#1{#1}\fi
\providecommand{\enquote}[1]{``#1''}
\expandafter\ifx\csname url\endcsname\relax
  \def\url#1{\texttt{#1}}\fi
\expandafter\ifx\csname urlprefix\endcsname\relax\def\urlprefix{URL }\fi
\providecommand{\eprint}[2][]{\url{#2}}
\vspace*{-0.1in}

\bibitem[{Leonard} and {Filippenko}(2005)]{Leonard11b}
D.~C. {Leonard}, and A.~V. {Filippenko}  (2005), in ASP Conf. Ser. 342:
  1604-2004: Supernovae as Cosmological Lighthouses, ed. M. Turatto, S.
  Benetti, L. Zampieri, \& W. Shea (San Francisco: ASP), 330-336.

\bibitem[{Smartt}(2009)]{Smartt09a}
S.~J. {Smartt}, \emph{\araa} \textbf{47}, 63--106 (2009).

\bibitem[{Nordhaus} et~al.(2010)]{Nordhaus10}
J.~{Nordhaus}, A. {Burrows}, A. {Almgren} and J. {Bell},
  \emph{\apj} \textbf{720}, 694--703 (2010).

\bibitem[{H{\" o}flich} et~al.(1996)]{Hoflich96}
P.~{H{\" o}flich}, J.~C. {Wheeler}, D.~C. {Hines}, and S.~R. {Trammell},
  \emph{\apj} \textbf{459}, 307--321 (1996).

\bibitem[{Leonard} et~al.(2001)]{Leonard3}
D.~C. {Leonard}, A.~V. {Filippenko}, D.~R. {Ardila}, and M.~S. {Brotherton},
  \emph{\apj} \textbf{553}, 861--885 (2001).

\bibitem[{Leonard} and {Filippenko}(2001)]{Leonard4}
D.~C. {Leonard}, and A.~V. {Filippenko}, \emph{\pasp} \textbf{113}, 920--936
  (2001).

\bibitem[{Wang} and {Wheeler}(2008)]{Wang08}
L.~{Wang}, and J.~C. {Wheeler}, \emph{\araa} \textbf{46}, 433--474 (2008).

\bibitem[{Leonard} et~al.(2006)]{Leonard13}
D.~C. {Leonard}, et al., \emph{\nat} \textbf{440}, 505--507 (2006).

\bibitem[{H{\" o}flich} (1991)]{Hoflich91}
P.~{H{\" o}flich},  \emph{\aap} \textbf{246}, 481--489 (1991).

\bibitem[{Chornock} et~al.(2010)]{Chornock10}
R.~{Chornock}, A.~V. {Filippenko}, W.~{Li}, and J.~M. {Silverman}, \emph{\apj}
  \textbf{713}, 1363--1375 (2010).

\bibitem[{Leonard} et~al.(2009)]{Leonard09}
D.~C. {Leonard}, A.~V. {Filippenko}, M.~{Ganeshalingam}, W.~{Li}, B.~{Swift},
  and T.~R. {Diamond}, \emph{Bulletin of the American Astronomical Society}
  \textbf{41}, 466 (2009).

\bibitem[{Dessart} and {Hillier}(2011)]{Dessart11}
L.~{Dessart}, and D.~J. {Hillier}, \emph{\mnras}  (2011), in press (ArXiv
  e-prints 1104.5346).

\bibitem[{Dessart} and {Hillier}(2011)]{Dessart11b}
L.~{Dessart}, and D.~J. {Hillier}, \emph{\mnras} \textbf{410}, 1739--1760 (2011).

\bibitem[{Maguire} et~al.(2010)]{Maguire10}
K.~{Maguire}, et al., \emph{\mnras}, \textbf{404}, 981--1004 (2011).

\bibitem[{Van Dyk} et~al.(2011)]{Vandyk11}
S.~D. {Van Dyk}, et al., \emph{\aj} (2011), submitted (ArXiv
  e-prints 1011.5873).

\bibitem[{Chugai}(2007)]{Chugai06}
N.~N. {Chugai}, \emph{Astronomy Letters}, \textbf{32}, 739--746 (2006).


\end{thebibliography}

\end{document}